\documentstyle[12pt]{article}
\def\setb@se#1{\baselineskip=#1 \normalbaselineskip=#1}
\setb@se{14pt}
\newcommand{\be}{\begin{equation}}
\newcommand{\ee}{\end{equation}}

\begin{document}

\begin{titlepage}
\begin{flushright}

ZU-TH 3/95

February 1995

hep-th/9502045
\end{flushright}

\

\vspace{20 mm}
\begin{center}
\Huge
The number of sphaleron  instabilities  of  the  Bartnik-McKinnon
solitons and non-Abelian black holes
\vspace{5 mm}
\end{center}

\begin{center}
{\bf  M.S. Volkov,
\footnote{On  leave  from  Physical--Technical
Institute of the  Academy  of  Sciences  of  Russia,  Kazan  420029,
Russia}
O. Brodbeck,
G. Lavrelashvili,
\footnote{On  leave of absence from Tbilisi Mathematical
Institute, 380093 Tbilisi, Georgia}
and N. Straumann}

\vspace{5 mm}
Institut    f\"ur    Theoretische    Physik    der    Universit\"at
Z\"urich-Irchel, Winterthurerstrasse 190, CH-8057 Z\"urich,

Switzerland

\end{center}
\vspace{20 mm}
\begin{center}
{\large Abstract}
\end{center}
\vspace{10mm}

It is proven that there are precisely $n$ odd-parity  sphaleron-like
unstable modes of the $n$-th Bartnik-McKinnon soliton and the $n$-th
non-abelian black hole solution of  the  Einstein-Yang-Mills  theory
for the gauge group $SU(2)$.

\end{titlepage}
\newpage

\section{Introduction}

It  is  by  now  well  known  that  Bartnik-McKinnon  (BK)  solitons
\cite{BK} and non-abelian  black  holes  \cite{BH}  of  the  $SU(2)$
Einstein-Yang-Mills (EYM) theory are very fragile objects. A  linear
perturbation analysis \cite{SZ}  showed  that  these  solutions  are
unstable  and  non-linear  numerical  studies  \cite{ZS},   \cite{Z}
revealed that the instability is quite dramatic.

The  unstable  modes   found   in   \cite{SZ}-\cite{Z}   belong   to
perturbations within the original BK  ansatz.  This
type of instabilities will  be  referred  to  as  ``gravitational'',
because they have no flat spacetime analogues. The  numerical  results
for the few lowest members of  the  solution  families  led  to  the
suspicion that the $n$-th EYM (soliton or black hole)  solution  has
exactly $n$ such unstable gravitational modes, a conjecture which is
still not proven.

Beside  these  even-parity  modes  there  is  a  second   class   of
exponentially growing modes, with the opposite parity, which we call
``sphaleron-like''  instabilities,  because  they  have  a  similar
origin as for the sphaleron solution of the Yang-Mills-Higgs  system
in flat spacetime. Heuristically,  this  type  of  instabilities  is
related to the existence of non-contractible loops in  configuration
space, passing through the equilibrium solution (at  least  for  $n$
odd) \cite{M}. This fact suggests that the latter are some  kind
of saddle points. For odd $n$'s there are also paths  connecting
homotopically  distinct  vacua,  differing  in  their   Chern-Simons
number, which pass through the equilibrium  solutions  \cite{sphal},
\cite{V}. It has been shown analytically that there exists at  least
one sphaleron-like unstable mode for each member of  the  BK  family
and this remains true for ``generic'' solitons belonging to  general
gauge groups \cite{BS1}, \cite{BS2}. The same has been shown for the
$SU(2)$ EYM black holes \cite{odd} and ``generic'' non-abelian black
holes for any compact  gauge  group  \cite{BS2}.  Numerical  studies
\cite{LM} for the few lowest members of the BK family again suggests
that there are also exactly $n$ sphaleron-like instabilities. It  is
the main purpose of the present paper to prove this conjecture, both
for BK solitons and non-abelian black holes.

Our demonstration is based on the study of some  ``dual''  pairs  of
Schr\"odinger equations, whose partners are formally related in  the
same manner as in supersymmetric  quantum  mechanics.  Thanks  to  a
residual gauge  group,  it  is  possible  to  find  the  zero-energy
solutions in sufficiently explicit form to read off  the  number  of
their nodes and hence, for an appropriate choice  of  the  pair,  the
number of bound states, i.e.,  unstable  modes.  We  find  it  quite
remarkable that this can be deduced, in spite of the fact  that  the
BK solutions  and  their  black  hole  counterparts  are  not  known
analytically.  (Rigorous  existence  proof  have   been   given   in
\cite{SM}, \cite{S}, \cite{BFM}).

\section{Basic equations}

The basic equations for spherically symmetric $SU(2)$ EYM fields, as
well as for spherical perturbations of the EYM  solitons  and  black
holes, have been derived  before  (even  for  general  gauge  groups
\cite{BS1}). We follow the notations in Ref.\cite{guys},  where  the
time-dependent spherically-symmetric gravitational and gauge  fields
are parameterized as follows.

For the metric we use Schwarzschild coordinates
\be
ds^{2} = -N S^{2}dt^{2} +N^{-1} dr^{2} +
r^{2}(d\vartheta^{2} + \sin^{2}\vartheta
d\varphi^{2}),                                     \label{1}
\ee
and the $SU(2)$ gauge potential is represented as
\be
A = a_{0}\tau_{r}\ dt + a_{1}\tau_{r}\ dr  +(w-1)
(\tau_{\varphi}d\vartheta-\tau_{\vartheta}\sin\vartheta ~d\varphi)+
\tilde{w}
(\tau_{\vartheta}d\vartheta+\tau_{\varphi}\sin\vartheta ~d\varphi).
                                                          \label{2}
\ee
The metric coefficients  $N$,  $S$  and  gauge  amplitudes  $a_{0}$,
$a_{1}$,  $w$,  $\tilde{w}$  are  functions  of  $r$  and  $t$,  and
$\tau_{r}$,   $\tau_{\vartheta}$,   $\tau_{\varphi}$   denote    the
spherical   generators   of   $SU(2)$,    normalized    such    that
$[\tau_{r},\tau_{\vartheta}]=\tau_{\varphi}$, etc.

The coupled EYM field equations, expressed in terms of the functions
in (\ref{1}) and (\ref{2}), can be  found  in  \cite{guys}.  In  the
static case, with  $a_{0}=a_{1}=\tilde{w}=0$,  these  equations  are
known  to  admit  soliton  and  black  hole   solutions   \cite{BK},
\cite{BH}, characterized  by  the  number  of  nodes  of  the  gauge
amplitude $w(r)$. Here we repeat  only  the  resulting  perturbation
equations for these  equilibrium  solutions  \cite{SZ},  \cite{odd},
\cite{guys}.

Since  the   parity   transformation,   $\vartheta\rightarrow\pi   -
\vartheta$,  $\varphi\rightarrow\varphi   +\pi$,   is   a   symmetry
operation, the even-parity perturbations $(\delta N,\delta  S,\delta
w)$    and    the    odd-parity    modes    $(\delta    a_{0},\delta
a_{1},\delta\tilde{w})$  decouple. The perturbation equation for the
even-parity (``gravitational'') modes can be brought into  the  form
of a $P$-wave Schr\"odinger equation \cite{SZ}
\be
-\eta''+\left(S^{2}N\frac{3w^{2}-1}{r^{2}}+
2\left(\frac{S'}{S}\right)'\right)\eta
=\omega^{2}\eta,                                        \label{3}
\ee
where $\delta w=\eta(\rho)\exp(i\omega t)$, and a prime
always denotes differentiation with respect
to the radial coordinate $\rho$ defined by
\be
\frac{d\rho}{dr}=\frac{1}{NS}                          \label{4}
\ee
($\rho(0)=0$ for solitons and $\rho=-\infty$ at the horizon
for black holes).

For the odd-parity (``sphaleron'') modes we choose the
temporal gauge, $\delta a_{0}=0$, and adopt the following
notations for the perturbation amplitudes
$\delta a_{1}$,  $\delta\tilde{w}$:
\be
\delta a_{1}=\alpha(\rho)\exp(i\omega t),\ \ \
\delta\tilde{w}=\xi(\rho)\exp(i\omega t).              \label{5}
\ee
For $\alpha$ and $\xi$ we obtain the coupled system,
consisting of the second order equation
\be
-\xi''+
\frac{N S^{2}}{r^{2}}(w^{2}-1)\xi+
NSw'\alpha+(NSw\alpha)'
=\omega^{2}\xi,                                         \label{6}
\ee
and the first order equation
\be
\frac{NS^{2}}{r^{2}}w^{2}\alpha+
\frac{S}{r^{2}}(w'\xi -
w\xi')=\frac{1}{2}\omega^{2}\alpha.                    \label{7}
\ee
In addition, the Gauss constraint gives
\be
\omega
\left\{ \left(\frac{r^{2}}{S}\alpha\right) '-
2w\xi\right\} =0.                                         \label{8}
\ee
The following observation simplifies the further discussion of  this
type of perturbations considerably. If we  multiply  (\ref{6})  with
$w$ and take the background Yang-Mills equation
\be
w''=\frac{NS^{2}}{r^{2}}w(w^{2}-1)                \label{8:1}
\ee
into account,
the left hand side of (\ref{6}) becomes a total derivative
of the left hand side of (\ref{7}), multiplied by $r^{2}/S$.
Introducing the quantities
\be
\chi=\frac{r^{2}}{2S}\alpha,\ \ \
\gamma^{2}=\frac{2NS^{2}}{r^{2}},                   \label{9}
\ee
we find from (\ref{6})-(\ref{8}):
\be
\left( w^{2}\gamma^{2}\chi+w'\xi-w\xi'\right)'=
\omega^{2}w\xi,                                   \label{10}
\ee
\be
 w^{2}\gamma^{2}\chi+w'\xi-w\xi'=\omega^{2}\chi,  \label{11}
\ee
\be
\omega (\chi'-w\xi)=0.                            \label{12}
\ee
When $\omega=0$, there is a residual $U(1)$ gauge symmetry
of these equations:
\be
\xi\rightarrow\xi+w\Omega,\ \ \ \
\chi\rightarrow\chi+\Omega'/\gamma^{2},            \label{13}
\ee
where $\Omega$ is a function of $\rho$. This will be of
crucial importance.

\section{The number of odd-parity sphaleron instabilities}

In  this  section   we   prove   that   the   number   of   negative
$(\omega^{2}<0)$ modes of the system (\ref{10})-(\ref{12})  for  the
$n$-th BK solution, as well as for the $n$-th  EYM  black  hole,  is
exactly $n$. We call  this  type  of  instabilities  sphaleron-like,
because of their similarity with the flat spacetime Yang-Mills-Higgs
sphaleron instabilities \cite{sphal}.

For $\omega\neq 0$, Eq.(\ref{10}) is a consequence of (\ref{11})
and (\ref{12}), i.e.,
\be
w\xi'-w'\xi=(w^{2}\gamma^{2}-\omega^{2})\chi,             \label{14}
\ee
\be
\chi'=w\xi.                                               \label{15}
\ee
As a technical tool we keep equation (\ref{15}) also for  $\omega=0$
and call it {\sl strong Gauss constraint}; it plays the  role  of  a
gauge fixing condition. It  suffices  to  determine  the  number  of
$\omega^{2}<0$ modes of the system (\ref{14}), (\ref{15}). (The role
of  the  Gauss  constraint  is   discussed   more   extensively   in
\cite{BS2}).

We begin our analysis by eliminating $\xi$, with the result
\be
-\phi''+
\left(\frac{1}{2}
\gamma^{2}(w^{2}+1)+2\left(\frac{w'}{w}\right)^{2}\right)
\phi
=\omega^{2}\phi,                                        \label{16}
\ee
where $\phi=\chi/w$. For a smooth function $w$  without  zeros  this
would be a regular Schr\"odinger equation  with  a  purely  positive
spectrum $(\omega^{2}\geq 0)$. This  suggests  that  the  number  of
unstable modes of the $n$-th equilibrium solution is indeed  related
to the number $n$ of nodes of $w$.

Since the differential equation (\ref{16}) has singularities at  the
node positions of  $w$,  we  try  to  pass  to  a  ``dual''  regular
Schr\"odinger equation, for which the number of bound states can  be
read off.  How  to  achieve  this  is  suggested  by  the  following
considerations.

First, we note that Eq.(\ref{16}) has obvious $\omega=0$
solutions which are just pure gauge modes (see (\ref{13})),
\be
\phi_{0}=\Omega'/w\gamma^{2},                          \label{17}
\ee
satisfying the strong Gauss constraint (\ref{15}) with
$\xi=w\Omega$:
\be
\left(\frac{\Omega'}{\gamma^{2}}\right)'=w^{2}\Omega. \label{18}
\ee
Any such solution allows us to factorize the differential
operator in (\ref{16}):
\be
-\frac{d^{2}}{d\rho^{2}}+
\left(\frac{1}{2}
\gamma^{2}(w^{2}+1)+2\left(\frac{w'}{w}\right)^{2}\right)=
-\frac{d^{2}}{d\rho^{2}}+\frac{\phi_{0}''}{\phi_{0}}
=Q^{+}Q^{-},                                            \label{19}
\ee
with
\be
Q^{\pm}=\mp\frac{d}{d\rho}-\frac{\phi_{0}'}{\phi_{0}}. \label{20}
\ee
Using (\ref{18}) we find
\be
Q^{\pm}=\mp\frac{d}{d\rho}+\frac{w'}{w}+w^{2}Z,         \label{21}
\ee
where
\be
Z=-\Omega\Lambda,\ \ \ \Lambda=\gamma^{2}/\Omega'.   \label{22}
\ee
For later use we note that (\ref{18}) is equivalent to
\be
\Lambda'/\Lambda=w^{2}Z,                             \label{23}
\ee
and  thus $Z$ satisfies the nonlinear differential equation
\be
Z'=w^{2}Z^{2}-\gamma^{2}.                             \label{24}
\ee

With  a standard reduction, the most general solution of the
second order linear equation (\ref{18}) is
\be
\Omega=c_{2}\tilde{\Lambda}+\tilde{\Omega}
\left(c_{1}+c_{2}\int_{0}^{\rho}w^{2}
\tilde{\Lambda}^{2}d\rho\right),                    \label{25}
\ee
where $\tilde{\Omega}$ is a special solution, $\tilde{\Lambda}=
\gamma^{2}/\tilde{\Omega}'$, and $c_{1}$, $c_{2}$ are real
constants. This gives immediately for $Z$ in (\ref{22})
\be
Z=-\tilde{\Omega}\tilde{\Lambda}-
\frac{\tilde{\Lambda}^{2}}
{c_{1}/c_{2}+{\displaystyle\int_{0}^{\rho}}w^{2}
\tilde{\Lambda}^{2}d\rho}.                           \label{26}
\ee

Since the differential equation (\ref{16}) is, for any $\Omega$
in (\ref{25}), identical to
\be
Q^{+}Q^{-}\phi=\omega^{2}\phi,                       \label{27}
\ee
we are invited to pass from $\phi$ to
\be
\psi=Q^{-}\phi.                                      \label{28}
\ee
For $\omega\neq 0$ this has the unique inverse
\be
\phi=\frac{1}{\omega^{2}}Q^{+}\psi                   \label{29}
\ee
and by applying $Q^{-}$ we obtain the ``dual'' eigenvalue
equation
\be
Q^{-}Q^{+}\psi=\omega^{2}\psi.                       \label{30}
\ee
The differential operator on the left is
\be
Q^{-}Q^{+}=-\frac{d^{2}}{d\rho^{2}}+U,              \label{31}
\ee
with
\be
U=\frac{1}{2}\gamma^{2}(3w^{2}-1)+
2\left( w^{2}Z\right)'.                               \label{32}
\ee
In this potential the nodes of $w$ are no more dangerous
and we have the chance that (\ref{30}) becomes a
``regular'' Schr\"odinger equation if $\Omega$ in (\ref{25})
is chosen appropriately.

Before addressing this problem, we determine the zero-energy
solutions of (\ref{30}). These satisfy $Q^{+}\psi_{0}=0$, i.e.,
\be
-\frac{\psi_{0}'}{\psi_{0}}=\frac{w'}{w}+w^{2}Z=
\frac{\phi_{0}'}{\phi_{0}},                         \label{33}
\ee
thus
\be
\psi_{0}=\frac{1}{\phi_{0}}=w\Lambda=
w\frac{\tilde{\Lambda}}
{c_{1}+c_{2}{\displaystyle\int_{0}^{\rho}}w^{2}
\tilde{\Lambda}^{2}d\rho}.                           \label{34}
\ee
An alternative form is obtained from (\ref{23}), (\ref{34}):
\be
\psi_{0}=w\exp\left(\int_{\rho_{0}}^{\rho}w^{2}Zd\rho\right).
                                                     \label{36}
\ee
This looks good, because the factor $w$ would count  the  number  of
nodes of this zero-energy solution if the second factor on the right
hand side could be chosen to be everywhere regular.

A detailed discussion of the function  $Z$  in  (\ref{26}),  whereby
$\tilde{\Omega}$ is  a  special  solution  of  the  linear  equation
(\ref{18}), will be given elsewhere. Here we only state that one can
choose $\tilde{\Omega}$ such that
\be
\tilde{\Omega}= \frac{const}{r^{2}}+O(1)\ \
{\rm as}\ r\rightarrow 0,\ \ \ \ \ \ \ \ \ \
\tilde{\Omega}= \frac{1}{r^{2}}+O(\frac{1}{r^{3}})
\ \ {\rm as}\ r\rightarrow\infty                  \label{37}
\ee
for the BK solitons, and
\be
\tilde{\Omega}= x+O(x^{2})\ \
{\rm as}\ x\rightarrow 0,\ \ \ \ \ \
\tilde{\Omega}= const~x+O(1)
\ \ {\rm as}\ x\rightarrow\infty                  \label{38}
\ee
for the EYM black holes, where  $x=(r-r_{h})/r_{h}$,  $r_{h}$  being
the Schwarzschild coordinate of the horizon.

For the regular solutions, one obtains from  (\ref{37}),  (\ref{26})
(with an appropriate constant $c_{2}/c_{1}$ in (\ref{26})) that  $Z$
exists everywhere on the half-line with the following behavior  near
$\rho=0,\infty$:
\be
Z= \frac{1}{\rho}+O(\rho)\ \ {\rm as}\   \rho\rightarrow 0,
\ \ \ \ \
Z= -\frac{2}{\rho^{2}}+O(\frac{ln(\rho)}{\rho^{3}})
\ \ {\rm as}\  \rho\rightarrow\infty.                \label{39}
\ee
This assures that the potential  $U$  in  (\ref{32})  is  everywhere
bounded. Indeed, the first term for $U$ behaves as $2/\rho^{2}$  for
$\rho\rightarrow 0$, and this singularity is exactly canceled by the
second term in $U$ involving the function $Z$.

In summary, the ``dual'' equation (\ref{30}) can be chosen such that
it becomes an $S$-wave Schr\"odinger  equation  with  an  everywhere
bounded potential. In  addition  (\ref{39})  assures  the  following
behavior for the zero-energy solution $\psi_{0}$:
\be
\psi_{0}\sim\rho\ \ {\rm as}\ \ \rho\rightarrow 0,
\ \ \ \ \ \ \ \ \
\psi_{0}\sim\frac{1}{\rho^{2}}
\ \ {\rm as}\ \ \rho\rightarrow\infty,  \label{40}
\ee
which implies  that  $\psi_{0}\in  L^{2}[0,\infty)$.  A  well-known
theorem ensures then that the number of nodes of  $\psi_{0}$,  which
is $n$ for the $n$-th BK soliton solution, is  also  equal  the  the
number of bound states of (\ref{30}), and thus equal to  the  number
of sphaleron-like instabilities.

Similarly,  for  black  holes,  the  special  solution   (\ref{38}),
together with an appropriate choice of $c_{2}/c_{1}$ in  (\ref{26}),
ensure that $Z$ has the following behavior near $\rho=\pm\infty$:
\be
Z= -\frac{1}{f^{2}(r_{h})\rho}+O(\frac{1}{\rho^{2}})\ \ {\rm as}\ \
\rho\rightarrow -\infty,\
\ \ \ \ \
Z\rightarrow -\frac{2}{\rho^{2}}+O(\frac{ln(\rho)}{\rho^{3}})
\ \ {\rm as}\ \
\rho\rightarrow\infty.                                  \label{41}
\ee
This yields
\be
\psi_{0}\sim\frac{1}{\rho}\ \ {\rm as}\ \ \rho\rightarrow -\infty,
\ \ \ \ \ \  \ \
\psi_{0}\sim\frac{1}{\rho^{2}}\ \ {\rm as}\ \  \rho\rightarrow\infty,
\ee
thus $\psi_{0}\in  L^{2}(-\infty,\infty)$,  implying  that  for  the
$n$-th EYM black hole (independent of its event horizon size)  there
exist precisely $n$ unstable odd-parity modes.

\section{Concluding remarks}

Looking back at the proof, the following elements have been crucial.
Thanks to the residual gauge symmetry (\ref{13}) we know zero energy
solutions of the original differential equation  (\ref{16})  and  we
can thus factorize the differential operator.  We  can  choose  this
factorization  such  that  the  ``supersymmetric  partner''   is   a
well-behaved  Schr\"odinger  equation  which  has   a   normalizable
zero-energy solution. For this solution, we can determine the number
of zeros and hence read off the number of bound states.

Unfortunately, because $\eta$ in Eq.(\ref{3})
remains invariant under  the  residual
gauge transformation, this method does not work for the  even-parity
gravitational instabilities, whose number is presumably the same.

\section*{Acknowledgments}

This  work was supported by the Swiss National Science Foundation
and by the Tomalla Foundation.

\end{document}